\documentclass[twocolumn,showpacs,preprintnumbers,nofootinbib,amsmath,amssymb]{revtex4}
\input{psfig.sty}

\usepackage{dcolumn}
\usepackage{bm}

\begin{document}

\title{Lookback time as a test for brane cosmology}

\author{N. Pires$^1$} \email{npires@dfte.ufrn.br}

\author{Zong-Hong Zhu$^2$} \email{zhuzh@bnu.edu.cn}

\author{J. S. Alcaniz$^3$} \email{alcaniz@on.br}

\affiliation{$^1$Departamento de F\'{\i}sica, Universidade Federal do Rio Grande do Norte,
59072-970, Natal, Brasil}

\affiliation{$^2$Department of Astronomy, Beijing Normal University, Beijing 100875, China}

\affiliation{$^3$Departamento de Astronomia, Observat\'orio Nacional, 20921-400, Rio de
Janeiro, Brasil}

\date{\today}

\begin{abstract}

The observed late-time acceleration of the Universe may be the result of unknown physical processes involving either  modifications of gravitation theory or the existence of new fields in high energy physics. In the former case, such modifications are usually related to the possible existence of extra dimensions (which is also required by unification theories), giving rise to the so-called brane cosmology. In this paper we investigate the viability of this idea by considering a particular class of brane scenarios in which a large scale modification of gravity arises due to a gravitational \emph{leakage} into extra dimensions. To this end, differently from other recent analyses, we combine \emph{orthogonal} age and distance measurements at intermediary and high redshifts. We use observations of the lookback time to galaxy clusters, indirect estimates of the age of the Universe from the most recent Large-Scale Structure (LSS) and Cosmic Microwave Background (CMB) data, along with the recent detection of the baryon acoustic oscillations at $z = 0.35$. In agreement with other recent analyses we show that, although compatible with these age and distance measurements, a spatially closed scenario is largely favoured by the  current observational data. By restricting our analysis to a spatially flat universe, we also find that the standard $\Lambda$CDM model is favoured over the particular braneworld scenario here investigated.

\end{abstract}

\maketitle

\section{Introduction}

Since the close of the past century, an increasing number of observational results have
transformed radically our view and understanding of the Universe. From these observations
a consistent picture has emerged indicating that we live in a \emph{dark pressure} (or
\emph{dark energy}) dominated universe, whose the current rate of expansion is an
increasing function of time. If confirmed, this extra component, or rather, its
gravitational effects, will be the very first observational piece of evidence for new
physics beyond the domain of the standard model of Particle Physics. Such a expectation
has, in turn, given rise to many speculations on the fundamental nature of the dark
component which dominates the current dynamics of the Universe (see, e.g,
\cite{review} for a recent review).

Among several alternatives to this dark energy problem, a very interesting one suggests
modifications in gravity instead of any adjustment to the energy content of the Universe.
Examples of modified gravity models include, among others, scenarios with higher order
curvature-invariant modifications of the Einstein-Hilbert action \cite{carroll} and
braneworld (BW) cosmologies \cite{braneS}. While in the first case the \emph{natural}
matter dilution in the expanding universe is avoided by adding high order terms to the
gravitational sector of the theory, in the latter example our 4-dimensional Universe is
thought of as a surface or a brane embedded into a higher dimensional bulk space-time on
which only gravity could propagate. If such extra dimensions were  unambiguosly detected
from cosmological observations, BW models could not only give a solution to the so-called
dark energy problem but also provide a natural explanation for the huge difference
between the two fundamental energy scales in nature, namely, the electroweak and Planck
scales [$M_{Pl}/m_{EW} \sim 10^{16}$] (see also \cite{randall}).

Recently, many attempts to observationally detect or distinguish brane effects from the
usual dark energy physics have been discussed in the literature.   In Ref. \cite{ss}, for
instance, Sahni and Shtanov investigated a class of BW models which admit a wider range
of possibilities for the dark pressure than do the usual dark energy scenarios. As shown
by these authors (also in Ref. \cite{ss1}), a new and interesting feature of this class
of models is that the acceleration of the Universe may be a transient
phenomena\footnote{Some dark energy scenarios also admit this transient acceleration.
See, e.g., \cite{alcstef}}, which cannot be achieved in the context of our current
standard scenario, i.e., the $\Lambda$CDM model but could reconcile the supernova
evidence for an accelerating universe with the requirements of string/M-theory \cite{fis}
(see also \cite{brane} for other theoretical and observational aspects of BW models).

Another particularly interesting BW model is the one proposed by Dvali {\it{et al.}}
\cite{dgp}, which is widely refered to as DGP model. This scenario describes a
self-accelerating 5-dimensional BW model with a noncompact, infinite-volume extra
dimension in which the dynamics of gravitational interaction is governed by a competition
between a 4-dimensional Ricci scalar term, induced on the brane, and an ordinary
5-dimensional Einstein-Hilbert action. For scales below a crossover radius $r_c$ (where
the induced 4-dimensional Ricci scalar dominates), the gravitational force experienced by
two punctual sources is the usual 4-dimensional $1/r^{2}$ force whereas for distance
scales larger than $r_c$ the gravitational force follows the 5-dimensional $1/r^{3}$
behavior\footnote{As is well known, according to Gauss' law the gravitational force falls
off as $r^{1 - S}$ where $S$ is the number of spatial dimensions.}. The theoretical
consistency of the model, and in particular of its self-accelerating solution, is still a
matter of debate in the current literature (see, e.g., \cite{luty}). From the
observational viewpoint, however, DGP models have been successfully tested in many of
their predictions, ranging from local gravity to cosmological observations
\cite{deff1,alc1,alc2,zhuacl,lue} (see also \cite{lue05} for a recent  review on the DGP
phenomenology).

In this paper we are particularly interested in testing the viability of DGP scenarios
from cosmological  time measurements, i.e., observations of lookback time to galaxy
clusters at intermediary and high redshifts and  recent estimates of the total age of the
Universe. In order to build up the lookback time sample we use the age estimates of
$\sim$ 160 galaxy clusters at six redshifts lying in the interval $0.10\leq z \leq 1.27$,
as compiled by Capozziello {\it et al.} \cite{capo}, and assume the total expanding age
of the Universe to be $t^{obs}_o = 14.8 \pm 0.7$ Gyr, as obtained by MacTavish {\it et
al.} \cite{agesdss} from an analysis involving the most recent Large-Scale Structure (LSS)
and Cosmic Microwave Background (CMB) data. We also show that the constraints from
lookback time observations on the parameters of the model are orthogonal to those
obtained from recent measurements of the baryon acoustic oscillations at $z = 0.35$
\cite{bao}, providing very restrictive bounds on DGP cosmologies.  In agreement with
other independent analyses, it is shown that  spatially closed DGP models are largely
favoured by the current data.

\begin{figure*}
 \centerline{\psfig{figure=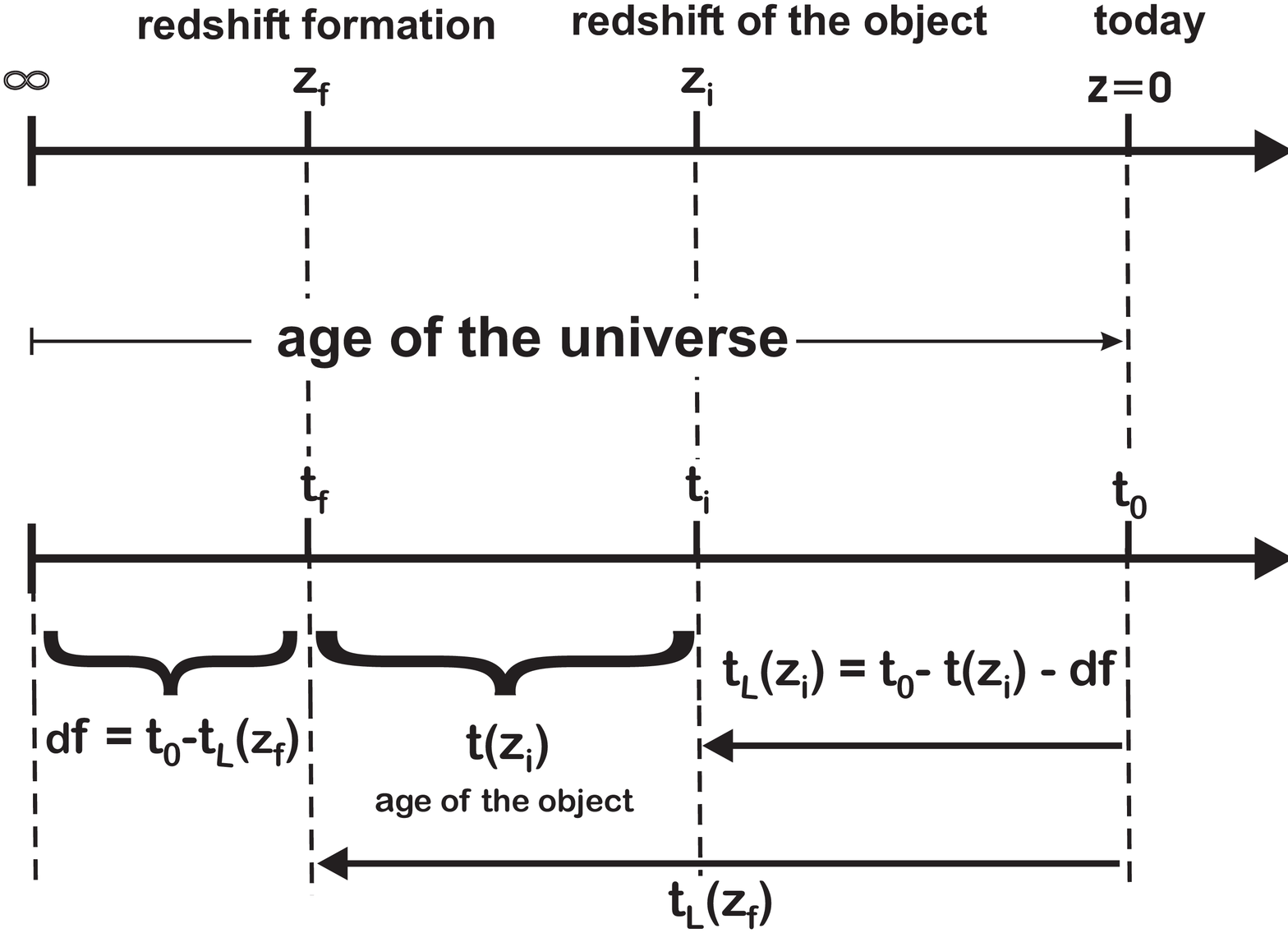,width=6.7truein,height=2.9truein,angle=0}} 
{\hskip 0.1in}
 \caption{Lookback time, age of the Universe, and related quantities.}
\end{figure*}

\section{Basics of DGP Models}

In DGP models, the modified Friedmann's equation due to the presence of an
infinite-volume extra dimension reads \cite{deff1}
\begin{equation}
\left[\sqrt{\frac{\rho}{3M_{pl}^{2}} + \frac{1}{4r_{c}^{2}}} +
\frac{1}{2r_{c}}\right]^{2} = H^{2} + \frac{k}{R(t)^{2}},
\end{equation}
where $H$ and  $\rho$ are, respectively, the Hubble parameter and the energy density of
the cosmic fluid (which we will assume to be composed only of nonrelativistic particles),
$k = 0, \pm 1$ is the spatial curvature parameter,  $M_{pl}$ is the Planck mass and $r_c
= M_{pl}^{2}/2M_{5}^{3}$ ($M_5$ is the 5-dimensional reduced  Planck mass) is the
crossover scale defining the gravitational interaction among particles located on the
brane.  Note that for values of  $\rho/3M_{pl}^{2} >> 1/r_{c}^{2}$, DGP and the standard
cold dark matter (SCDM)  models are analogous so that the cosmological evolution for the
early stages of the Universe is exactly the same in both scenarios.

Equation (1) can be rewritten as
 \begin{eqnarray}
 {\cal{H}}^2  = \Omega_k (1 + z)^{2} + \left[\sqrt{\Omega_{\rm{r_c}}} +
\sqrt{\Omega_{\rm{r_c}} + \Omega_{\rm{m}}(1 + z)^{3}}\right]^{2},
 \end{eqnarray}
where ${\cal{H}} \equiv {H(z)}/{H_o}$ (${H_o}$ is the current value of the Hubble
parameter),  $\Omega_{\rm{m}}$ and $\Omega_k$ stand for  the matter and curvature density
parameters, respectively, and
\begin{equation} \label{rc}
\Omega_{\rm{r_c}} = 1/4r_c^{2}H_o^{2},
\end{equation}
is the density parameter associated to the crossover radius $r_c$. Note still that the
general  normalization condition in DGP scenarios is given by
\begin{equation}
\Omega_k + \left[\sqrt{\Omega_{\rm{r_c}}} + \sqrt{\Omega_{\rm{r_c}} +
\Omega_{\rm{m}}}\right]^{2} = 1,
\end{equation}
 while for a flat universe ($\Omega_k = 0$), it  reduces to $\Omega_{\rm{r_c}} = (1 -
\Omega_{\rm{m}})^{2}/4$. As noticed in Ref. \cite{deff1}, the above described cosmology
can be exactly reproduced by the standard one plus an additional dark energy component
with a time-dependent  equation of state parameter
\begin{equation}
\omega^{eff}(z) = 1/{\cal{G}}(z, \Omega_{\rm{m}},\Omega_{\rm{r_c}}) - 1,
\end{equation}
 where
\begin{eqnarray}
 {\cal{G}}(z, \Omega_{\rm{m}},\Omega_{\rm{r_c}}) = \sqrt{4\Omega_{\rm{r_c}}/
 \Omega_{\rm{m}}{x'}^{-3} +  4}(\sqrt{\Omega_{\rm{r_c}}/\Omega_{\rm{m}}{x'}^{-3}} + \nonumber \\
 + \sqrt{\Omega_{\rm{r_c}}/\Omega_{\rm{m}}{x'}^{-3} + 1}), \quad \quad \quad \quad \quad
\end{eqnarray}
and ${x'} = (1 + z)^{-1}$.

As shown by several authors \cite{alc1,alc2,zhuacl,lue}, for values of the length scale $r_c
\simeq H^{-1}_o$ (see also Table I), the presence of an infinite-volume extra dimension
as described above leads to a late-time acceleration of the Universe, in agreement with
most of the current distance-based cosmological observations. In the next Section,
differently from the above analyses, we test the viability of these scenarios from time
measurements, i.e., measurements of lookback time to galaxy clusters and recent estimates
of the age of the Universe (see also \cite{alc2} for an analysis involving age estimates
of high-$z$ objects in the context of BW models).

\section{Confronting the model with lookback time data}

In his seminal \emph{Observational Tests for World Models}, Sandage \cite{sand} defines the lookback
time as the difference between the present age of the Universe ($t_o$)  and its age
($t_z$) when a particular light ray at redshift $z$ was emitted. For many years, this
interesting relation (which involves the main cosmological parameters) was used as a
cosmological probe only in a qualitative way due exclusively  to the lack of reliable age
data. Only more recently, with the advent of large telescopes and new technics, it was
possible to estimate with reasonable precision ages of high-$z$ objects, including
galaxies, quasars and galaxy clusters (see, for instance, \cite{ages,quasar}).

Here, in order to discuss age constraints on the BW scenarios presented above, we use the age estimates of $\sim$ 160 galaxy clusters at six redshifts distributed in the interval $0.10\leq z \leq 1.27$, as compiled by Capozziello {\it et al.} \cite{capo}. The clusters at intermediary redshifts (i.e., $z=0.60$, $0.70$ and $0.80$) have their age estimated from the color of their reddest galaxies, whereas for the two at lower redshifts and the high-$z$ galaxy clusters at $z = 1.24$ \cite{blake}, it was used the color scattering from a large sample of low redshifts SDSS clusters imaged \cite{andreon}  (see Sec. IV of  \cite{capo} for more details on the age sample). The total age of the Universe is assumed in our analysis to be  $t^{obs}_o = 14.8 \pm 0.7$ Gyr, as obtained by MacTavish {\it et al.} \cite{agesdss} from a joint analysis involving
recent LSS data (the matter power spectra from the 2dFGRS and SDSS redshift surveys) and results of the most recent CMB experiments (WMAP, DASI, VSA, ACBAR, MAXIMA, CBI and BOOMERANG).

\subsection{The Lookback Time Test}

The lookback time-redshift relation, as defined above, is given by
\begin{equation} \label{looktheo}
t_L(z;\mathbf{p}) = H^{-1}_o \int_o^z{\frac{dz'}{(1 + z'){{\cal{H}}(\mathbf{p})}}},
\end{equation}
where $H^{-1}_o = 9.78h^{-1}$ Gyr and $\mathbf{p}$ stands for the density parameters
$\Omega_{\rm{m}}$ and $\Omega_{\rm{r_c}}$ (throughout this paper we assume $h = 0.72 \pm
0.08$, as provided by the HST key project \cite{hst}). Now, following Capozziello {\it et
al.} \cite{capo}, let us consider an object (e.g., a galaxy, a quasar or a galaxy
cluster) at redshift $z_i$ whose the age $t(z_i)$ is defined as the difference between
the age of the Universe at $z_i$ and the one when the object was born (at its formation
redshift $z_F$),  i.e.,
\begin{eqnarray}
t(z_i)  =  {H^{-1}_o} \left[\int_{z_i}^{\infty}{\frac{dz'}{(1 +
z'){{\cal{H}}(\mathbf{p})}}} -  \int_{z_F}^{\infty}{\frac{dz'}{(1 +
z'){{\cal{H}}(\mathbf{p})}}}\right]
\end{eqnarray}
or, equivalently,
\begin{equation}
t(z_i)  = t_L(z_F) - t_L(z_i).
\end{equation}
(see the schema presented in Fig. 1).

From the above expressions (and Fig. 1), it is now straightforward to define the observed
lookback time to an object at $z_i$ as
\begin{eqnarray} \label{lookobs}
 t^{obs}_L(z_i) & = & t_L(z_F) - t(z_i)  \nonumber \\ & &
= [t^{obs}_o - t(z_i)] - [t^{obs}_o - t_L(z_F)] \nonumber \\ & &
= t^{obs}_o - t(z_i) - df,    
\end{eqnarray}
where $df$ stands for the \emph{incubation time} or \emph{delay factor}, which accounts
for our ignorance about  the amount of time since the beginning of the structure
formation in the Universe until the formation time ($t_f$) of the object.

\begin{figure*}
 \centerline{ 
 \psfig{figure=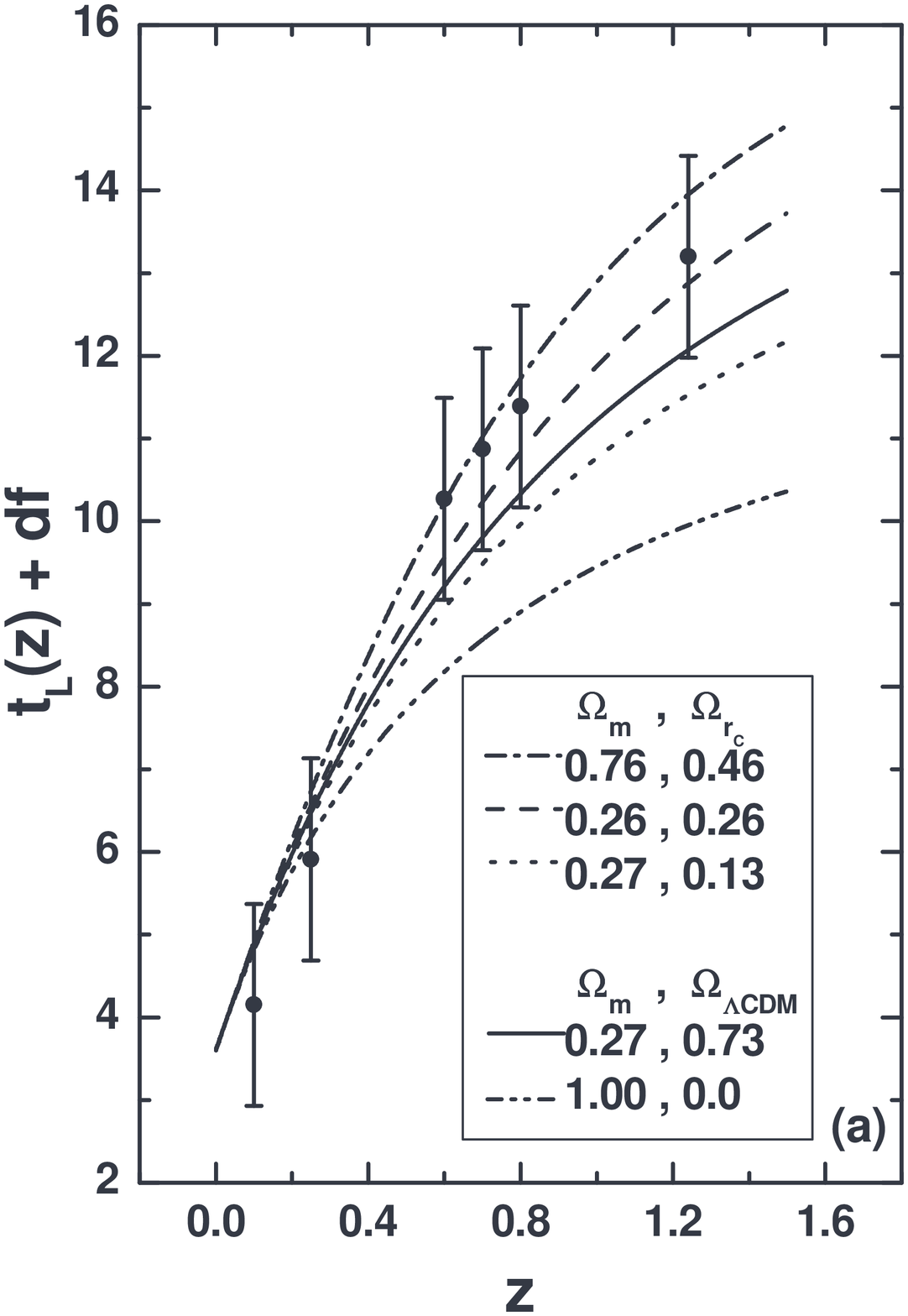,width=6.0truecm} \hskip 0.1truecm
 \psfig{figure=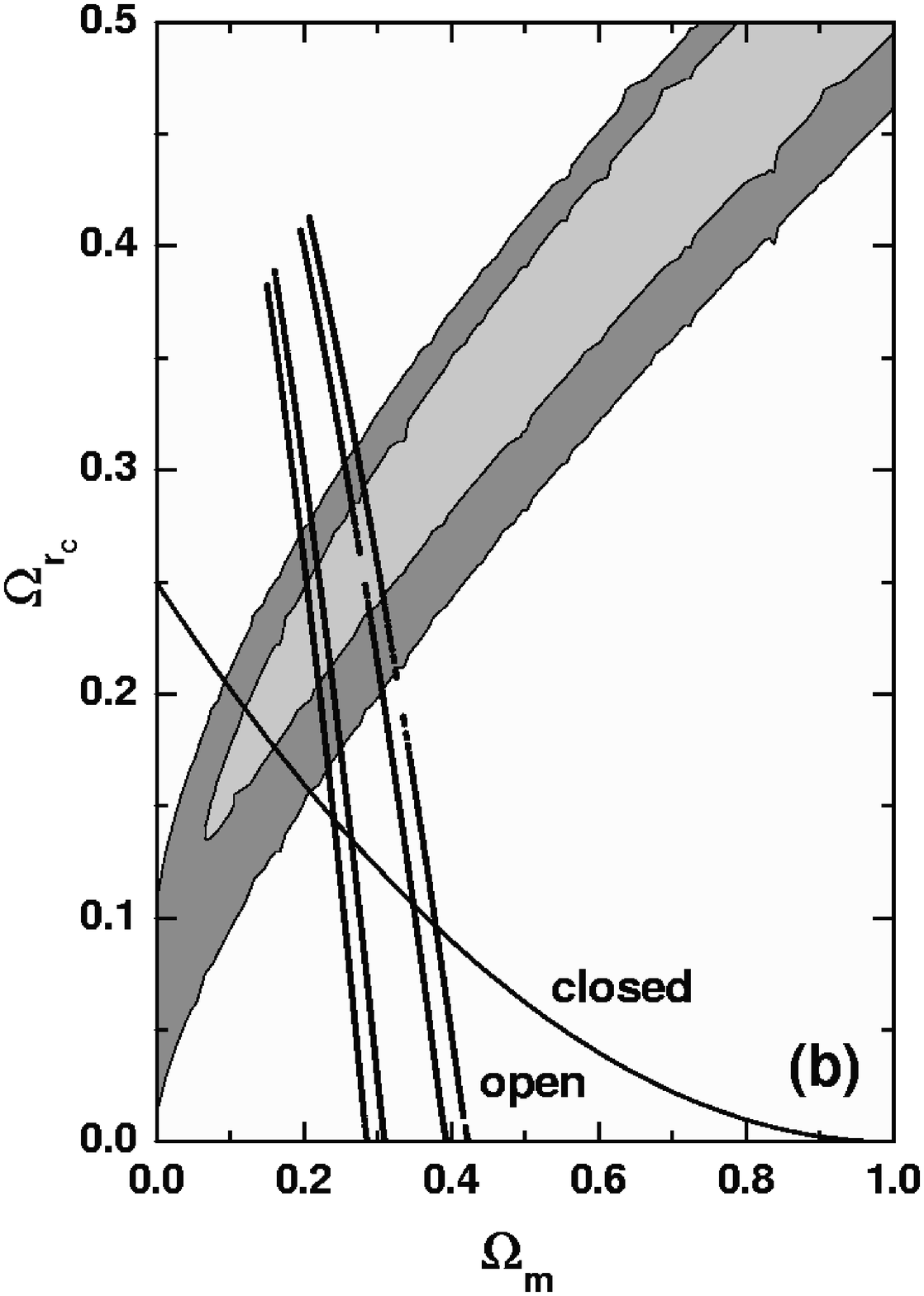,width=6.0truecm}\hskip 0.1truecm
 \psfig{figure=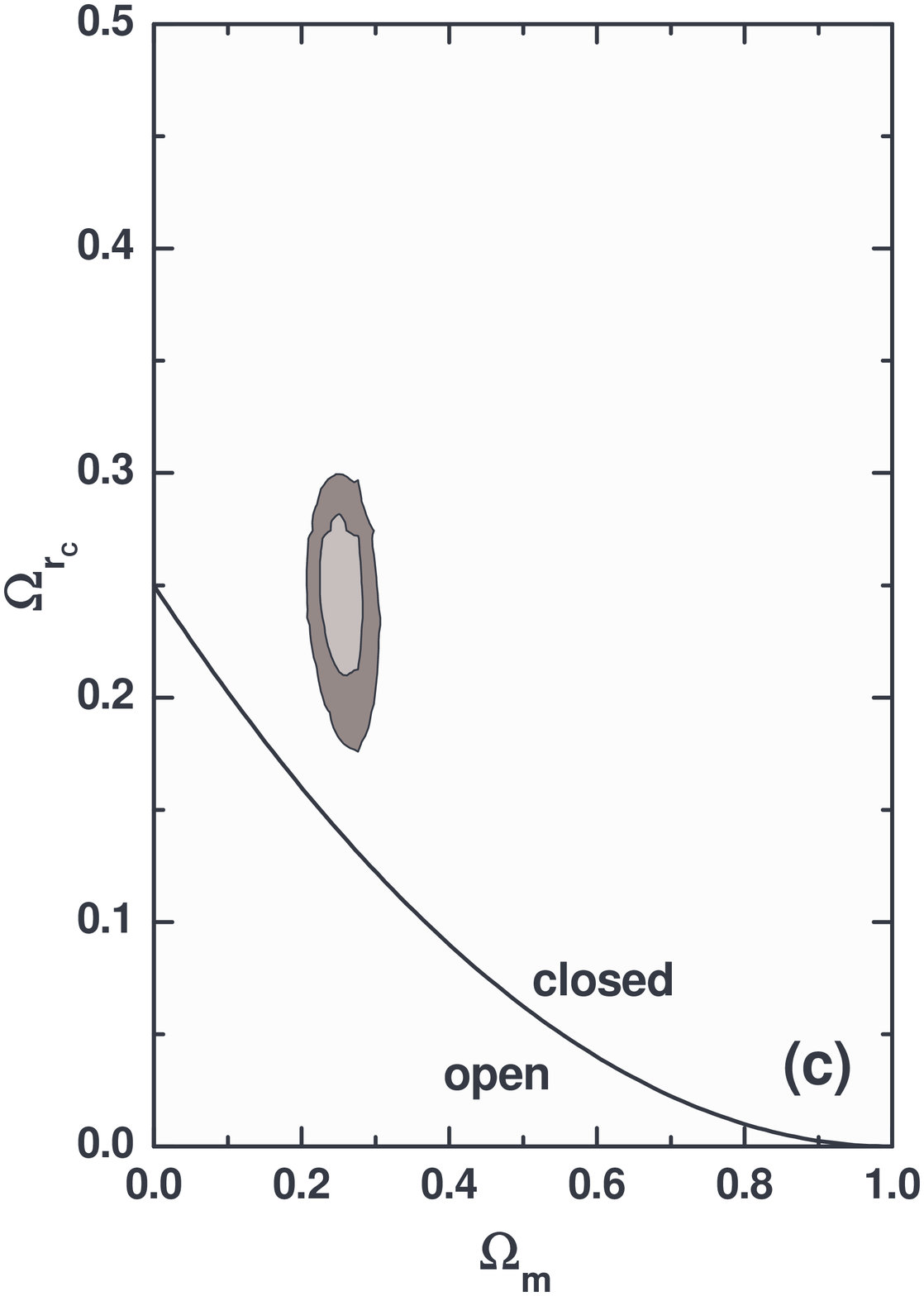,width=6.0truecm} 
 \hskip 0.1in}
\caption{The results of the lookback time and BAO analyses. {\it (a)} The lookback time relation as a function of the redshift  for values of the density parameters  $\Omega_m$ and $\Omega_{r_c}$ corresponding to the best-fits of the statistical analyses. To plot these curves, we have fixed $df = 3.6$ Gyr. {\it (b)} 68.3 \% and 95.4 \% confidence regions in the $\Omega_m - \Omega_{r_c}$ plane considering only the age of the galaxy clusters. Dotted lines stand for the corresponding $1\sigma -2\sigma$ regions arising from baryon oscillation peak. {\it (c)} The $1\sigma-2 \sigma$ confidence regions for the joint analysis of age of the clusters and BAO. Note that, in both cases, a spatially closed scenario is largely favoured.}
\end{figure*}

\subsection{Statistical Analysis}

In order to estimate the best fit to the set of parameters $\mathbf{p} \equiv \{ \Omega_{\rm{m}},
\Omega_{\rm{r_c}}\}$ we define the likelihood function
\begin{equation}
{\cal{L}}_{age} \propto \exp\left[-\chi_{age}^{2}(z;\mathbf{p},df)/2\right],
\end{equation}
with the $\chi_{age}^{2}$ function given by
\begin{eqnarray} \label{chi2}
\chi_{age}^{2} &  = & \sum_{i}{\frac{\left[t_L(z_i;\mathbf{p}) -
t^{obs}_L(z_i)\right]^{2}} {\sigma_i^{2} + \sigma_{t^{obs}_o}^{2}}} + \nonumber \\ & &
\quad \quad \quad \quad   + \frac{\left[t_o(\mathbf{p}) -
t^{obs}_o\right]^{2}}{\sigma_{t^{obs}_o}^{2}}.
\end{eqnarray}
In the above expression, $\sigma_i = 1$ Gyr is the uncertainty in the individual lookback
time to the i$^{\rm{th}}$ galaxy cluster of our sample while $\sigma_{t^{obs}_o} = 0.7$
Gyr stands for the uncertainty on the total expanding age of the Universe ($t^{obs}_o$).
As discussed in Ref. \cite{capo}, two important aspects concerning the above equations
should be emphasized at this point. First, that we have included a prior on the total age
of the Universe -- the second term of Eq. (\ref{chi2}). The reason for this additional
term is the well-known fact that the evolution of the age of the Universe with redshift
($dt_{\rm{U}}/dz$) may differ considerably  from scenario to scenario, which means that
it is possible that cosmological models that are able to explain age estimates of
high-$z$ objects may not be compatible with the total expanding age at $z = 0$ (and
vice-versa) -- see \cite{quasar} for an example. The second aspect concerns the delay
factor $df$. Note that while the observed lookback time $t^{obs}_L(z_i)$  [Eq.
(\ref{lookobs})] depends explicitely on $df$ its theoretical value $t_L(z_i;\mathbf{p})$ [Eq.
(\ref{looktheo})] does not.  Moreover, it is also worth mentioning that in principle it must be different for each object in the sample. Here, however, the delay factor $df$  is assumed as a ``nuisance" parameter, so that we marginalize it over the interval [0, 20] Gyr.

In Fig. 2a we show the binned data of the lookback time plotted as a function of redshift
for a fixed value of $df = 3.6$ Gyr and some selected values of the density parameters $\Omega_{\rm{m}}$ and
$\Omega_{\rm{r_c}}$. For the sake of comparison, the Einstein-de Sitter model and standard $\Lambda$CDM prediction
(with $\Omega_{\rm{m}} = 0.27$ and $\Omega_{\rm{\Lambda}} = 0.73$) are also shown. Figure 2b shows the first results of our statistical analysis. Confidence regions
($68.3\%$ and  $95.4\%$) in the $\Omega_{\rm{m}} - \Omega_{\rm{r_c}}$ space are displayed
by considering the lookback time measurements discussed above. Note that  the area
corresponding to the confidence intervals is not very restrictive, a fact that is expected and understood in terms
of the conservative uncertainty assumed  ($\sigma_i = 1$ Gyr) for the  individual
lookback time (see Fig. 2a)\footnote{As well emphasized in Ref. \cite{capo},  the value
of $\sigma_i = 1$ Gyr is a conservative assumption, in the sense that if the uncertainty
on the galaxy cluster age were so large, then there also should be observed a large
scatter in the color - magnitude relation for the reddest cluster galaxies, which is not.
However, to take into account other systematic uncertainties as, for instance, those
relative to  the evolutionary model, we consider that the value assumed, although
conservative, is well applicable.}. The best-fit parameters for this analysis are
$\Omega_{\rm{m}} = 0.85$ and $\Omega_{\rm{r_c}} = 0.49$, which corresponds to a spatially
closed scenario, in agreement with other independent analyses. (See, e.g.,
\cite{zhuacl} and references therein). If, however, the Gaussian prior on the matter density parameter
$\Omega_{\rm{m}} = 0.27 \pm 0.04$ (as given by WMAP results \cite{wmap})  is assumed, the
new best-fit model is found at $\Omega_{\rm{m}} = 0.27$ and $\Omega_{\rm{r_c}} = 0.26$.
By restricting the analysis to the flat case, we note that the data favour a lower value of the matter density parameter, i.e., $\Omega_{\rm{m}} = 0.14 \pm 0.04$ ($\Omega_{\rm{r_c}} = 0.18$) at 95.4\% (c.l.). Also, if we compare the flat cases of DGP and $\Lambda$CDM scenarios, we find that the former is disfavored, with $\chi^2_{min}(\rm{DGP}) \simeq 1.28 \chi^2_{min}(\rm{\Lambda CDM})$.

\subsection{Joint Statistics with BAO}

As well known,  the acoustic peaks in the cosmic microwave background (CMB) anisotropy power spectrum is an efficient way for determining cosmological parameters (e.g., \cite{wmap}).  Because the acoustic oscillations in the relativistic plasma of the early universe will also be imprinted on to the late-time power spectrum of the non-relativistic matter \cite{peeblesyu}, the acoustic signatures in the large-scale clustering of galaxies yield additional tests for cosmology. In particular, the characteristic and reasonably sharp length scale measured at a wide range of redshifts provides an estimate of the distance-redshift relation, which is a geometric complement to the usual luminosity-distance from type Ia supernove \cite{bao}. Using  a large spectroscopic sample of 46,748 luminous, red galaxies (LRGs) covering 3816 square degrees out to a  redshift of z=0.47 from the Sloan Digital Sky Suvey, Eisenstein et al. \cite{bao} have successfully found the peaks, described by the ${\cal{A}}$-parameter, which is independent of cosmological models, i.e., 
\begin{eqnarray}
{\cal{A}} \equiv {\Omega_m^{1/2} \over {{\cal{H}}(z_{\rm{*}})}^{1/3}}\left[\frac{1}{z_{\rm{*}}\sqrt{|\Omega_k|}}{\cal{F}} \left(\sqrt{|\Omega_k|}\Gamma(z_*)\right)\right]^{2/3} \\  = 0.469 \pm 0.017, \nonumber
\end{eqnarray}
where $z_{\rm{*}} = 0.35$ is the redshift at which the acoustic scale has been measured,
$\Gamma(z_*) \equiv \int_0^{z_*}{{dz}/{{\cal{H}}(z_{\rm{*}})}}$ is the dimensionless comoving distance to $z_*$,  
and  the function
${\cal{F}}$ is defined by one of the following forms: ${\cal{F}}(x) = {\sinh}(x)$, $x$,
and ${\sin}(x)$, respectively, for open, flat and closed geometries (we refer the reader to Ref. \cite{bao} for more details on BAO physics).

The dotted lines in Fig. 2b represent  the constraints from this measurement on the
parametric space $\Omega_{\rm{m}} - \Omega_{\rm{r_c}}$ (see also \cite{ariel}). The
important point to be noted in Fig. 2b is that lookback time data and the BAO prior
provide orthogonal statistics in the plane $\Omega_m - \Omega_{\rm{r_c}}$. This,
therefore, suggests that possible degeneracies between these parameters may be broken by
combining these two kinds of observations in a joint statistical analysis. The results of
such an analysis are shown in Fig. 2c. Note that  the parameter space now is considerably
reduced relative to the former analysis, with the best-fit model occurring at
$\Omega_{\rm{m}} = 0.25$ and $\Omega_{\rm{r_c}} = 0.25$.  Such a model corresponds to an
accelerating universe with $q_o \simeq -0.83$, a total expanding age $t_o \simeq
10.5h^{-1}$ Gyr, and a transition redshift of the order of $z_{\rm{T}} \simeq 1.0$. At 95.4\% c.l., we also obtain  $0.23 \leq \Omega_{\rm{m}} \leq 0.29$ and $0.24 \leq \Omega_{\rm{r_c}} \leq  0.28$.

From Eq. (\ref{rc}), we also note that the joint best-fit value for $\Omega_{\rm{r_c}}$ ($0.25$) leads to an estimate of the crossover scale $r_c$ in terms of the
Hubble radius $H_o^{-1}$, i.e.,
\begin{equation}
r_c = 1.0 H_o^{-1}.
\end{equation}
If now we restrict our analysis to the flat case ($\Omega_k = 0$), we find $\Omega_{\rm{r_c}} = 0.13$ or, equivalently, $\Omega_{\rm{m}} = 0.27$, i.e., in good agreement with current $\Omega_{\rm{m}}$ estimates from CMB data \cite{wmap}. This particular
value of  $\Omega_{\rm{r_c}}$ corresponds to a crossover distance between 4-dimensional
and 5-dimensional gravities of the order of $r_c \simeq 1.38 H_o^{-1}$. In Table I we
summarize the main estimates for $r_c$ obtained in this paper.

\begin{table}
\caption{Best-fit values for $\Omega_{\rm{m}}$, $\Omega_{\rm{r_c}}$ and $r_{c}$}
\begin{ruledtabular}
\begin{tabular}{lcrl}
Test& $\Omega_{\rm{m}}$  & $\Omega_{\rm{r_c}}$  &$r_{c}$\footnote{in units of
$H_o^{-1}$}\\ \hline \hline \\ 
Lookback Time & 0.98  & 0.54 & 0.68 \\ 
Lookback Time + $\Omega_{\rm{m}}$ & 0.27 & 0.26 & 0.98 \\ 
Lookback Time + BAO & 0.25 & 0.25 & 1.0\\
\hline \\ & {flat case}& &\\ \hline \\ Lookback Time & 0.14  & 0.18 & 1.17 \\ Lookback
Time + BAO & 0.27 & 0.13 & 1.38\\ \hline  \\
\end{tabular}
\end{ruledtabular}
\end{table}

\section{Conclusion}

Alternative cosmologies from brane world model provide a possible mechanism for the present acceleration of the universe congruously suggested by various cosmological observations. In this paper we have focused our attention on one of these scenarios, the so-called DGP model, in which the acceleration is attributed to gravitational \emph{leakage} into extra dimensions. We have explored the constraints on the parameter space of the DGP scenario from measurements of lookback time to galaxy clusters and the age of the universe.  It is shown that the lookback time observations provide complementary and interesting constraints on the parameters of the model, though there is a degeneracy between $\Omega_{\rm{m}}$ and $\Omega_{r_{c}}$. When we further combined the recent measurements of the baryon acoustic oscillations at $z = 0.35$ found in the Sloan Digital Sky Survey data, a very stringent constraint on both $\Omega_{\rm{m}}$ and $\Omega_{r_{c}}$ is obtained. The resulting parameters range as $0.23 \leq \Omega_{\rm{m}} \leq 0.29$ and $0.24 \leq \Omega_{\rm{r_c}} \leq  0.28$ at 95.4\% confidence level, indicating a closed universe, which is consistent with other independent analyses. If, however, a flat universe is considered a priori (as usually done in the context of dark energy models), we find $\Omega_{\rm{r_c}} = 0.13$ or, equivalently, $\Omega_{\rm{m}} = 0.27$, which is in good agreement with current estimates of the matter density parameter from CMB data \cite{wmap}.
    
Finally, it is also important to emphasize that although the current lookback time data do not provide very restrictive bounds on the density parameters $\Omega_m$ and $\Omega_{r_c}$ of the DGP model, the method discussed in Sec. III (and also in Refs. \cite{capo,ages}), along with new and more precise age measurements of high-$z$ objects will certainly provide a new and complemetary tool to test the reality of the current cosmic acceleration as well as to distingush among the many alternative world models. The present work, therefore, highlights the cosmological interest in the observational search for old collapsed objects at low, intermediary and high redshifts.

\begin{acknowledgments}
The authors are very grateful to V.F. Cardone, M. Fairbairn, A. Goobar and G.S. Fran\c{c}a for valuable
discussions. NP is supported by PRONEX/CNPq/FAPERN. Z-HZ is supported by the National Natural Science Foundation of China, under Grant No. 10533010, and by SRF, ROCS, SEM of China. JSA is supported by CNPq (Brazilian Agency) under Grants No. 307860/2004-3 and 475835/2004-2 and by Funda\c{c}\~ao de Amparo \`a Pesquisa do Estado do Rio de Janeiro (FAPERJ) No. E-26/171.251/2004.

\end{acknowledgments}


\end{document}